\documentclass[twoside]{article}
\usepackage{qic,epsfig}
\usepackage{amsmath}
\usepackage{amssymb}
\usepackage{graphicx}
\usepackage{hyperref}
\usepackage{tabto}
\usepackage{caption}
\usepackage[latin1]{inputenc}

\usepackage{aas_macros}

\usepackage{tikz}
\input{Qcircuit}

\begin{document}
\setlength{\textheight}{8.0truein}    

\runninghead{Quantum algorithmic differentiation}
            {G. Colucci and F. Giacosa }

\normalsize\textlineskip
\thispagestyle{empty}
\setcounter{page}{80}

\copyrightheading{21}{1\&2}{2021}{0080-0094}

\vspace*{0.88truein}

\alphfootnote

\fpage{80}

\centerline{\bf QUANTUM ALGORITHMIC DIFFERENTIATION}
\vspace*{0.37truein}
\centerline{\footnotesize
GIUSEPPE COLUCCI\footnote{info.quantumquants@gmail.com}}
\vspace*{0.015truein}
\centerline{\footnotesize\it Quantum Quants}
\baselineskip=10pt
\centerline{\footnotesize\it Rotterdam, 3035 AW, The Netherlands}
\vspace*{0.015truein}
\centerline{\footnotesize\it Institute of Physics, 
Jan Kochanowski University, ul.
Uniwersytecka 7}
\baselineskip=10pt
\centerline{\footnotesize\it Kielce, 25-406, Poland}
\vspace*{0.015truein}
\centerline{\footnotesize\it de Volksbank N.V., Croeselaan 1}
\baselineskip=10pt
\centerline{\footnotesize\it Utrecht, 3521 BJ, The Netherlands}
\vspace*{10pt}
\centerline{\footnotesize 
FRANCESCO GIACOSA}
\vspace*{0.015truein}
\centerline{\footnotesize\it Institute of Physics, 
Jan Kochanowski University, ul.
Uniwersytecka 7}
\baselineskip=10pt
\centerline{\footnotesize\it Kielce, 25-406, Poland}
\vspace*{0.015truein}
\centerline{\footnotesize\it Institute for Theoretical Physics, Goethe University,
Max-von-Laue-Str.\ 1}
\baselineskip=10pt
\centerline{\footnotesize\it Frankfurt am Main, 60438, Germany}
\vspace*{0.225truein}
\publisher{July 10, 2020}{December 24, 2020}

\vspace*{0.21truein}

\abstracts{
In this work we present an algorithm to perform algorithmic differentiation in the context of quantum computing. 
We present two versions of the algorithm, one which is fully quantum and one which employees a classical step (hybrid approach). 
Since the implementation of elementary functions is
already possible on quantum computers, the scheme that we propose can be easily applied. 
Moreover, since some steps (such as the CNOT operator) can (or will be) faster on a quantum computer than on a classical one, our procedure may ultimately
demonstrate that quantum algorithmic differentiation has an
advantage relative to its classical counterpart.}{}{}

\vspace*{10pt}

\vspace*{3pt}
\communicate{S~Braunstein~\&~K~Moelmer}

\vspace*{1pt}\textlineskip    

\section{Introduction}
\label{sec-introduction}
Algorithmic differentiation (also known as \textit{automatic} or \textit{computational} differentiation, hereafter AD) \cite{1982ZaMM...62Q.355G, Neidinger2010, Naumann2012} has gained particular attention in the last years due to its practical use in finance, in particular in the context of pricing financial derivatives \cite{GreeksAD, Homescu2011}, as well as in data science and machine learning \cite{2015arXiv150205767G}. 

In finance, AD is used to compute sensitivities (or Greeks) of the price of financial instruments with respect to the underlying drivers both accurately (to machine precision) and efficiently \cite{nagGreeks,Henrard2017,GreeksAD}. The advantage of AD w.r.t. the standard numerical differentiation is particularly evident when dealing with complex financial instruments for which numerical pricing methods are traditionally used, e.g., Monte Carlo. In fact, the calculation of price sensitivities with numerical differentiation involves perturbing the underlying price drivers, repeating simulations and getting finite-difference approximations. The computational effort thus becomes very significant when the financial instrument has a large number of drivers, as is typically the case for the sophisticated models for which Monte Carlo simulations are used in practice.

In data science, a particular case of algorithmic differentiation, known as \textit{backpropagation}, is used in Artificial Neural Network training to calculate gradients of the loss function with respect to the weights of the network (\cite{2015arXiv150205767G} and refs. therein). With the advance of deep learning, the need of increased efficiency and precision in the calculation of the gradients arose due to the large number of layers used to build up the networks.

Currently, several attempts have been made to introduce the concept and framework of differentiation to the quantum computing world \cite{2005PhRvL..95e0501J, 2019PhRvA..99c2331S, 2019arXiv190608728P, 2019npjQI...5..113O}. In fact, the progress made in optimization, simulations \cite{2007CMaPh.270..359B, 2018PhRvA..98b2321R}, financial applications \cite{2019arXiv190405803M, 2019arXiv190808040K}, quantum chemistry \cite{2019arXiv190608728P}, and machine learning algorithms \cite{2013arXiv1307.0411L,2015ConPh..56..172S,2015arXiv151202900A,2017Natur.549..195B} on a quantum computer requires the calculation of derivatives.

Until now, however, to the best of our knowledge no algorithm for AD has been proposed in the context of quantum computation. Nevertheless, the advances in scientific computing algorithms for quantum computers \cite{1996PhRvA..54.1034B, 2000quant.ph..1106F, 2008JPhCS.128a2013A, 2013NJPh...15a3021C, 2014arXiv1406.2040W, 2015arXiv151108253B, 2018arXiv180503265H} allow for the definition of primitive functions which can be used for algorithmic differentiation. Inspired by this approach, we propose in this paper a framework to implement algorithmic differentiation on a quantum computer, the \textit{quantum algorithmic differentiation framework}. Two versions are proposed: a fully quantum version, with no need of a hybrid quantum-classical system, and one in which classical ancilla registers are used for a more practical implementation. The proposed framework allows for the calculation of a composite function and its derivative at a given point at \textit{quantum} machine precision. The precision of the quantum algorithmic differentiation depends on the number of available qubits to represent a number on the quantum computer. This may be somewhat demanding considering the limited resources available on currently existing quantum computing hardware. Nevertheless, this paper aims at extending the range of available algorithms for scientific computing on a quantum computer with an algorithm for automatic differentiation, which is important to quantum algorithms for optimization and machine learning.

The paper is structured as follows: in Section~\ref{sec-classical-algorithmic-differentiation} the classical algorithmic differentiation is introduced. Section~\ref{section-qad-composite-function} presents a translation of the classical framework for algorithmic differentiation to the quantum case. Here the operators and procedures to calculate a function and its derivative are introduced. Sec. \ref{section-qad-framework} formalizes the results from the previous section and outlines the algorithm to calculate the derivative of a composite function at a given point (held in a quantum register).
Finally in Section~\ref{sec-conclusions} we summarize our findings and outline possible future developments. 

\section{Algorithmic differentiation}
\label{sec-classical-algorithmic-differentiation}

Algorithmic differentiation (AD) \cite{Neidinger2010,Henrard2017} uses exact formulas and numerical values to compute the derivative of a composite function starting from known elementary (or primitive) functions. 
It involves no approximation error as in numerical differentiation. AD is a third alternative to symbolic and numerical differentiation, and is also called computational differentiation or automatic differentiation.

The basic idea behind AD is to calculate the derivative of each a primitive function in an iterative way, and store at each step a tuple containing the value of the function and of the derivative, $(val, der)$. For the sake of simplicity we shall call this tuple the \textit{valder} object. 

More precisely, AD corresponds to the computational implementation of the chain rule. In fact, by applying the chain rule repeatedly to primitive functions (and arithmetic expressions), derivatives of arbitrary order of composite functions can be computed automatically.

An example can be used to clarify it. Consider the function 
$$f(x) = x^2\cdot\sin(\log(x)).$$ 
Assume we want to calculate the derivative of this function at $x_0$. Let us consider the \textit{valder} object $(x_0,1)$ as input to our algorithm. 

The function $f(x)$ can be rewritten as:
\begin{eqnarray}
f(x) & = & u(x)\cdot h(\log(x))\nonumber\\
	& = & u(x)\cdot h(w(x))\nonumber.
\end{eqnarray}

We can now split and execute the calculation of the derivative of $f$ by calculating the derivatives of each component, $c(x)$. The algorithm is based on a series of iterations that take as input a \textit{valder} object $(v,d)$ and define a new \textit{valder} object as $(c(v), c^\prime(v)\cdot d)$, where $c(x)$ is the component we are looking at.

For the function $f(x)$ we have:
\begin{itemize}
	\item define the input \textit{valder} tuple $(x_0,1)$.
	\item for $u(x)$:
	\begin{itemize}
		\item calculate the derivative of $u$ at $x_0$ and define a new \textit{valder} object containing the value of the function $u(x_0)$ and derivative $u^\prime(x_0)\cdot1$ (where 1 is the input value from the input tuple).
		\item store the \textit{valder} tuple : $(u(x_0),u^\prime(x_0))=(x_0^2,2x_0)\equiv(val_1,der_1)$
	\end{itemize}
	\item for $h(w(x))$:
		\begin{itemize}
			\item calculate the values of $w(x_0)$ and $w^\prime(x_0)$. 
			\item define the \textit{valder} tuple $(w(x_0),w^\prime(x_0)\cdot1)=(\log(x_0),1/x_0)\equiv(y,d)$ to be the input for the next step.
			\item calculate the values of the function $h(y)$ and derivative $h^\prime(y)$.
			\item define the \textit{valder} tuple\\ $(h(y),h^\prime(y)\cdot d)=(\sin(\log x_0),\cos(\log x_0)/x_0)\equiv(val_2,der_2)$			
		\end{itemize}
	\item Finally calculate the derivative of the product from the two obtained tuples as:
	$$(val_1*val_2, val_1*der_2 + val_2*der_1).$$
\end{itemize}

This example shows that from the knowledge of primitive functions and their derivatives we can calculate the derivative of a complex function without the error associated to numerical differentiation.

It can be shown that for practical applications the following primitive functions are enough to handle most typical situations: $exp(u)$, $log(u)$, $sqrt(u)$, $sin(u)$, $cos(u)$, $tan(u)$, $asin(u)$, $atan(u)$. Further the following operations are needed: $plus(u, v)$, $minus(u)$, $times(u,v)$, $reciprocal(u)$.

For all of the above mentioned primitive functions an implementation in terms of quantum circuits already exists \cite{2013NJPh...15a3021C, 2015arXiv151108253B, 2018arXiv180503265H}. Therefore in the following we develop a method to perform AD on a quantum computer.

For the sake of simplicity we shall focus on the algorithmic differentiation in the \textit{forward mode}. See \cite{2015arXiv150205767G} for a review of the forward and reverse mode implementations of algorithmic differentiation. 

\section{Quantum algorithmic differentiation of a composite function}
\label{section-qad-composite-function}
Consider three quantum registers, where a generic register of $n$-qubits is represented by a state $\ket{s}$:
\begin{equation}\label{eq:number-encoded-state}
\ket{s} = \ket{s_m}\otimes\ket{s_{m-1}}\otimes...\ket{s_0}\otimes\ket{s_{-1}}\otimes...\otimes\ket{s_{m-n}}
\end{equation} 
The first $m$ (where $m<n$) terms are the integer part and the last $n-m$ terms are the fractional part of a number. In general, if $x_0$ is a number in $\mathbf{R}$, and $\ket{x_0}$ the $n$-qubit representation of the number with precision $\frac{1}{2^{n-m}}$.

In the following we shall perform calculations within a fixed-point arithmetic framework. In fact, implementing floating point arithmetic on a quantum computer is quite complicated and has a huge overhead for exponent alignment, mantissa shifting and renormalization. 
As stated in \cite{2015arXiv151108253B}, the benefit of using fixed-precision (both for quantum as well as for custom integrated circuits) are clear when dealing with resource utilization (e.g., this framework is used in low power applications such as mobile or embedded systems, where hardware support for floating-point operations is often lacking). Moreover, if the algorithms are designed to perform calculations that depend on addition and multiplication only (i.e., not division) then there is no overhead in tracking the location of the decimal point and error and cost estimates are easier to derive.

Assume we want to calculate the derivative of a function $f(x)$ at $x_0$.

The building blocks of the quantum algorithmic differentiation are the following:
\begin{itemize}
	\item \textit{Transfer} or \textit{Copy} operator, $\mathcal{C}$: a series of CNOT between the first and second quantum registers.
	\item \textit{Reset} procedure, $\mathcal{R}$: a procedure to reset a quantum register to 0 (as a value, therefore $\ket{s}=\ket{00...0}$). 
	\item $\mathcal{AD}(f)$: this operator calculates the value of a function and its derivative at a point $v$, respectively $f(v)$ and $f^\prime(v)$, for a set of known functions (e.g. $\sin, \log, \exp,$ etc.).
\end{itemize}

In the following we describe the outlined building blocks.

\subsection*{\textit{Transfer} or \textit{Copy} operator}
\label{subsection-transfer-operator}

The transfer (or copy) operator is an extension of the controlled-Not (CNOT) operator to the tensor product of more than two qubits. Such operator acts as a copy, or transfer, of the controlled qubit onto the target qubit, with the condition that the latter state is set to $\ket{0}$ (or $\ket{00...0}$):
$$
\Qcircuit @C=1em @R=1em {
\lstick{\ket{a}} & \ctrl{4} & \qw      & \qw      & \rstick{\ket{a}} \qw \\
\lstick{\ket{b}} & \qw      & \ctrl{4} & \qw      & \rstick{\ket{b}} \qw \\
\lstick{...} & \qw & \qw & \qw & \qw & ... \\
\lstick{\ket{z}} & \qw      & \qw      & \ctrl{4} & \rstick{\ket{z}} \qw \\
\lstick{\ket{0}} & \targ    & \qw      & \qw      & \rstick{\ket{a}} \qw \\
\lstick{\ket{0}} & \qw      & \targ    & \qw      & \rstick{\ket{b}} \qw \\
\lstick{...} & \qw & \qw & \qw & \qw & ... \\
\lstick{\ket{0}} & \qw      & \qw      & \targ    & \rstick{\ket{z}} \qw 
}
$$
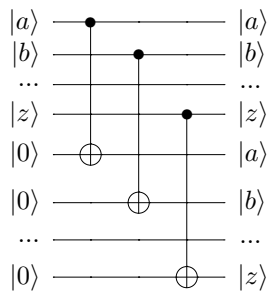
\captionof{figure}{\centering Quantum circuit representation of the Transfer operator $\mathcal{C}$.}

\tab\\where $a, b, ... ,z \in \{0,1\}$. Formally, this can be written as:
\begin{equation}
    \mathcal{C}\ket{a}\ket{b}...\ket{z}\ket{0}\ket{0}...\ket{0}=
    \ket{a}\ket{b}...\ket{z}\ket{a}\ket{b}...\ket{z}.
\end{equation}

In fact, in the proposed algorithm the control qubit will correspond to the value of a function at a point, while the target is initially set to $\ket{0...0}$. Therefore the goal of the transfer operator will be to copy the value of the function at a point onto the target register. Note that the second register should have at least the same number of qubits as the first register, to be able to represent the copy of the input value without incurring in precision loss.

Quite interestingly, there are ways to realize a fast CNOT obtained by employing Quantum Zeno dynamics \cite{fastcnot}, thus possibly rendering the whole procedure that we are about to describe more efficient and possibly even faster than its classical counterpart.

\subsection*{\textit{Reset} procedure}
\label{subsection-reset-procedure}

Our automatic differentiation algorithm needs to reset a certain quantum register to $\ket{00...0}$ in order to propagate the results of the intermediate calculations performed as it iterates applying the chain rule.

In general, one possibility would be to revert the unitary operations applied on the state. However, this could affect also states that we do not necessarily want to reset if operators across tensor space of several qubits are used. The following two possibilities are therefore considered:

\begin{itemize}
    \item Fully quantum implementation: prepare as many ancilla zero states ($\ket{0}$ or $\ket{00...0}$) as needed for the calculation. Whenever needed to reset the state, use a \textit{Swap gate} to swap the non-zero qubit with a qubit set to zero. 
    \item Hybrid implementation: for each qubit $q_i$ in the input state, consider a classical bit $c_i$. Do the following: 
    \begin{itemize}
        \item Measure the qubit $q_i$ on the standard Z-basis, storing the result on $c_i$.
        \item Apply to $q_i$ a bit-flip operator ($\sigma_X$). 
        \item If the corresponding classical bit, $c_i$, is zero apply again a bit-flip operator to the qubit $q_i$. Otherwise do nothing. 
        \item The qubit $q_i$ is then back in the zero state. 
    \end{itemize}
\end{itemize} 

For the sake of readability, in the following we will denote the Reset procedure by $\mathcal{R}$, without specifying which of the two implementations is used. Formally, we shall denote the action of the reset procedure in the operator form (non-unitary operator):
\begin{equation}
    \mathcal{R}\ket{a}\ket{b}...\ket{z}=
    \ket{0}\ket{0}...\ket{0}.
\end{equation}

\subsection*{$\mathcal{AD}(\cdot)$ operators}
\label{subsection-AD-operator}

The operator $\mathcal{AD}(f)$ is associated to a particular primitive function (e.g. $\sin$ or $\log$). Implementations for such functions are available in \cite{2013NJPh...15a3021C, 2015arXiv151108253B, 2018arXiv180503265H}. For each function $f(x)$, the operator takes as input three quantum registers, $\ket{a},\ket{b},\ket{c}$ and outputs the following three states: $\ket{f(a)},\ket{f^\prime(b)},\ket{f^\prime(b)*c}$. The operator $\mathcal{AD}(f)$ is built upon two operators: the first, $f\otimes f^\prime$, is represented by a multigate, in which one block acts on the first register, $\ket{a}$, to calculate $\ket{f(a)}$, and another block acts on $\ket{b}$ to calculate $\ket{f^\prime(b)}$; the second operator is a product operator, in which the product of the last two registers is calculated, $\ket{c\cdot f^\prime(b)}$. The definition of the latter operator can be found in \cite{2018arXiv180503265H}. 
In general, for a function $f(x)$ the operator $\mathcal{AD}(f)$ is shown in the figure below:
$$
\Qcircuit @C=2em @R=1em {
\lstick{\ket{a}} & \multigate{1}{f\otimes f^\prime}  & \qw & \qw                & \qw & \qw                   & \rstick{\ket{f(a)}} \qw \\
\lstick{\ket{b}} & \ghost{f\otimes f^\prime}         & \qw & \ket{f^\prime(b)}  &     & \multigate{1}{\times} & \rstick{\ket{f^\prime(b)}} \qw \\
\lstick{\ket{c}} & \qw                         & \qw & \qw                & \qw & \ghost{\times}        & \rstick{\ket{c\cdot f^\prime(b)}} \qw  \gategroup{1}{2}{3}{6}{1em}{--}
}
$$

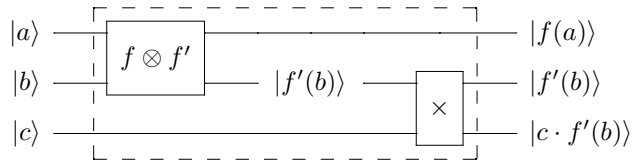
\captionof{figure}{\centering Circuit (operator) representing the module to calculate the derivative of the elementary function $f$.}

\tab\\In the quantum circuit above, the $\mathcal{AD}(f)$ operator is indicated by the dashed line.

Once this operator has been defined, the idea of algorithmic differentiation can then be applied as follows. Consider the composite function $g\circ f$. The first iteration of the algorithmic differentiation is to calculate the derivative of $f$ at $x_0$. It is then possible to calculate the derivative of $g\circ f$ in an iterative way. Let us consider the following input state:

\begin{equation}
	\ket{s} = \ket{x_0}\otimes\ket{0}\otimes\ket{1}
\end{equation}
where
\begin{itemize}
	\item $\ket{0} = \ket{00...0}$,
	\item $\ket{1} = \ket{0..010...0}$ is the ket which represents the integer number 1, i.e. the qubit set to 1 is the first digit on the left of the location of the decimal point, see (\ref{eq:number-encoded-state}). 
\end{itemize}

In analogy with the tuple used in Section~\ref{sec-classical-algorithmic-differentiation}, we call this state the \textit{valder state}.

Let us now apply the following algorithm:
\begin{itemize}
	\item apply the copy operator to the kets $\ket{x_0}$ and $\ket{0}$.
	\item apply the $\mathcal{AD}(f)$ operator to the first and second kets (both in state $\ket{x_0}$) and to the third ket in state $\ket{1}$.
\end{itemize}

The figure below shows the corresponding quantum circuit:
$$
\Qcircuit @C=1.5em @R=1em {
\lstick{\ket{x_0}} & \multigate{1}{\mathcal{C}}  & \qw & \ket{x_0} & & \multigate{2}{\mathcal{AD}(f)} & \qw  & \qw & \qw  & \qw      & \rstick{\ket{f(x_0)}} \qw \\
\lstick{\ket{0}} & \ghost{\mathcal{C}}         & \qw & \ket{x_0} & & \ghost{\mathcal{AD(f)}}        & \qw & \ket{f^\prime(x_0)}        & & \measure{\mbox{Reset}} & \rstick{\ket{0}} \qw \\
\lstick{\ket{1}} & \qw & \qw & \qw & \qw     & \ghost{\mathcal{AD(f)}}        & \qw & \qw & \qw & \qw      & \rstick{1\cdot\ket{f^\prime(x_0)}} \qw
}
$$\captionof{figure}{\centering Quantum circuit representing the sequence of modules to calculate the derivative of the elementary function $f$ at the point $x_0$.}

\tab\\So the output of the first iteration is the state:
\begin{eqnarray}
\ket{s} & = & \ket{f(x_0)}\otimes\ket{0}\otimes\ket{f^\prime(x_0)}\\
& = & \ket{val}\otimes\ket{0}\otimes\ket{der}
\end{eqnarray}
where for ease of notation we defined the kets $\ket{val}$ and $\ket{der}$.

Let us now consider the next iteration, where the only difference is that we have a function $g$ so we use the operator $\mathcal{AD}(g)$ and the input is the result of the previous iteration, $\ket{val}\otimes\ket{0}\otimes\ket{der}$:

\begin{itemize}
	\item apply the copy operator to the kets $\ket{val}$ and $\ket{0}$.
	\item apply the $\mathcal{AD}(g)$ operator (in this case associated to $g$) to the first and second kets (both in state $\ket{val}$) and to the third ket in state $\ket{der}$.
\end{itemize}

The figure below shows the corresponding quantum circuit:
$$
\Qcircuit @C=1.7em @R=1em {
\lstick{\ket{val}} & \multigate{1}{\mathcal{C}}  & \qw & \ket{val} & & \multigate{2}{\mathcal{AD}(g)} & \qw & \qw & \qw  & \qw      & \rstick{\ket{g(val)}} \qw \\
\lstick{\ket{0}} & \ghost{\mathcal{C}}         & \qw & \ket{val} & & \ghost{\mathcal{AD}(g)}        & \qw & \ket{g^\prime(val)}        & & \measure{\mbox{Reset}} & \rstick{\ket{0}} \qw \\
\lstick{\ket{der}} & \qw & \qw & \qw & \qw     & \ghost{\mathcal{AD}(g)}        & \qw & \qw & \qw & \qw      & \rstick{\ket{der\cdot g^\prime(val)}} \qw
\gategroup{1}{2}{3}{10}{3em}{--}
}
$$
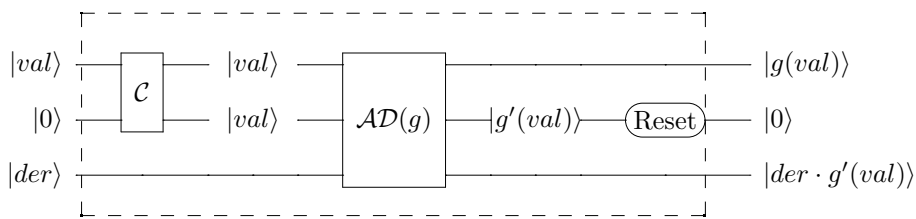
\captionof{figure}{\centering Circuit ($B_g$) representing the sequence of modules ($\mathcal{C}$, $\mathcal{AD}(\cdot)$ and $\mathcal{R}$) to iteratively calculate the derivative of a composite function.}

\tab\\
In this way we have calculated the derivative of the function $g \circ f$ at $x_0$.

One can extend this procedure to an arbitrary composition of functions. To this end, let us denote by $B_g$ the set of the operators (dashed box) in the previous circuit, where $B$ denotes the block and $g$ refers to the function $g(x)$. 

Similarly, if we intend to evaluate the derivative of the arbitrary composite function $f\circ g\circ...\circ h$, we have to apply the composition of blocks $B_{f},  B_{g},..., B_{h}$. The initial input on the left is
represented by the three-register state $\left\vert x_{0}\right\rangle \left\vert
0\right\rangle \left\vert 1\right\rangle $ and the output on the right is
\begin{equation}
\left\vert f(g(...h((x_{0}))\right\rangle \left\vert 0\right\rangle \left\vert
\left(  \frac{df}{dg}\right)  _{g_{0}}\cdot...\cdot\left(  \frac{dh}%
{dx}\right)  _{x_{0}}\right\rangle \text{ .}%
\end{equation}
The value in the third register of the output state is the derivative of the composite function that we were looking for. 

Note that at each intermediate step of the outlined procedure, the top and middle output registers may have different sizes or may hold values that are not in the same range. Therefore their sizes and forms (integer and fractional parts) need to be adjusted to account for the application of the Transfer operator at the next iteration (i.e. the second register must have at least the same number of qubits as the first register). 

Moreover, we assume we have quantum algorithms and circuits computing elementary functions. Since the derivative of an elementary function is an elementary function itself, the evaluation of the derivative of an arbitrary composite function is a straightforward (and systematic) application of our approach. 

\section{Quantum algorithmic differentiation: the framework}
\label{section-qad-framework}
In this section we merge the results from the previous two sections to define a quantum algorithmic differentiation ($q\mathcal{AD}$) framework.

Let us first define the functions needed for $q\mathcal{AD}$ in terms of a symbolic class:\\\\
\texttt{class QuantumAglorithmicDifferentiator()\\
	\indent attributes\\
	\indent\hspace{1cm}'valder' QuantumState $\ket{s} = \ket{v}\otimes\ket{0}\otimes\ket{d}$\\
	\indent methods\\
	\indent \hspace{1cm}\textit{arithmetic operations}\\
	\indent \hspace{1cm}ADplus($\ket{s1},\ket{s2}$)\\
	\indent \hspace{2cm}return $\ket{v1 + v2}\otimes\ket{0}\otimes\ket{d1 + d2}$\\
	\indent \hspace{1cm}ADminus($\ket{s}$)\\
	\indent \hspace{2cm} return $\ket{-v}\otimes\ket{0}\otimes\ket{-d}$\\
	\indent \hspace{1cm} ADtimes($\ket{s1}$,$\ket{s2}$)\\
	\indent \hspace{2cm} return $\ket{v1*v2}\otimes\ket{0}\otimes\ket{d1*v2 + v1*d2}$\\
	\indent \hspace{1cm} ADreciprocal($\ket{s}$)\\
	\indent \hspace{2cm} return $\ket{1/v}\otimes\ket{0}\otimes\ket{-1/v^2*d}$\\
	\indent \hspace{1cm}\textit{primitive functions}\\
	\indent \hspace{1cm} ADexp($\ket{s}$)\\
	\indent \hspace{2cm}return $\ket{\exp(v)}\otimes\ket{0}\otimes\ket{\exp(v)*d}$\\
	\indent \hspace{1cm} ADlog($\ket{s}$)\\
	\indent \hspace{2cm}return $\ket{\log(v)}\otimes\ket{0}\otimes\ket{1/v*d}$\\
	\indent \hspace{1cm} ADsqrt($\ket{s}$)\\
	\indent \hspace{2cm}return $\ket{\sqrt{v}}\otimes\ket{0}\otimes\ket{0.5/\sqrt{v}*d}$\\
	\indent \hspace{1cm} ADsin($\ket{s}$)\\
	\indent \hspace{2cm}return $\ket{\sin(v)}\otimes\ket{0}\otimes\ket{\cos(v)*d}$\\
	\indent \hspace{1cm} ADcos($\ket{s}$)\\
	\indent \hspace{2cm}return $\ket{\cos(v)}\otimes\ket{0}\otimes\ket{-\sin(v)*d}$\\
	\indent \hspace{1cm} ADtan($\ket{s}$)\\
	\indent \hspace{2cm}return $\ket{\tan(v)}\otimes\ket{0}\otimes\ket{1/\cos^2(v)*d}$\\
	\indent \hspace{1cm} ADarcsin($\ket{s}$)\\
	\indent \hspace{2cm}return $\ket{\arcsin(v)}\otimes\ket{0}\otimes\ket{1/\sqrt{1 - v*v}*d}$\\
	\indent \hspace{1cm} ADarctan($\ket{s}$)\\
	\indent \hspace{2cm}return $\ket{\arctan(v)}\otimes\ket{0}\otimes\ket{1/(1 + v*v)*d}$\\
}

The functions above defined are in fact operators acting on the quantum (\textit{valder}) states $\ket{s} = \ket{v}\otimes\ket{0}\otimes\ket{d}$. 
Note that for the sake of simplicity, we assume that the reset procedure $\mathcal{R}$ is always applied at the end of the application of the $\mathcal{AD}(\cdot)$ operators. 
Once these operators are implemented on a quantum computer, we can compute with any desired precision (subject only to the truncation error due to the number of allocated qubits, see Section~\ref{sec-note-error} for further details) the derivative of any combination of the functions with the following algorithm:

\tab

\noindent\fbox{
\parbox{\textwidth}{
\rule{\textwidth}{2pt}
\textbf{Algorithm $q\mathcal{AD}$($\ket{s}, n, m$)}\\
\rule{\textwidth}{2pt}
\indent \textbf{Require:} $\ket{s} = \ket{x_0}\otimes\ket{0}\otimes\ket{1}$ is a \textit{valder} state.\\
\begin{enumerate}
	\item Determine the components of the computational graph for the Algorithmic Differentiation.
	\item Calculate the number of qubits needed. 
	
	Originally the first and second registers have sizes sufficient to hold the input value $x_0$. The subsequent registers have sizes determined by the desired accuracy and the specifications of the quantum circuits used for the evaluation of the elementary functions appearing in the particular computational graph for Algorithmic Differentiation.
	
    \item For each component:
		\begin{enumerate}
			\item apply the transfer operator $\mathcal{C}$ between the first and second register.
			\item apply the component specific $\mathcal{AD}(\cdot)$ operator
			\item apply the reset procedure on the second register
			\item return the \textit{valder} object
		\end{enumerate}
	\item \textbf{If} arithmetic operations are involved, for each operation:
		\begin{enumerate}
			\item apply the component specific qAD operator
		\end{enumerate}
\end{enumerate}
}}

\subsection{An example: $f(x) = x\cdot\cos(\log x)$}
\label{subection-qad-framework-example}
The computational graph of the function $f(x) = x\cdot\cos(\log x)$ is shown below:
\begin{center}
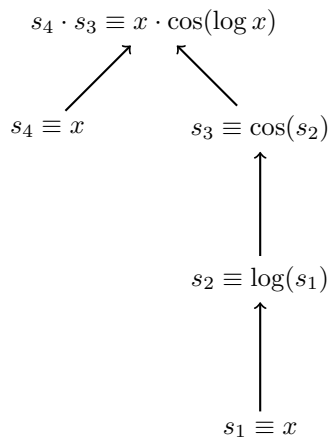

    \begin{tikzpicture}[->,auto,node distance=2cm, thick]
      \tikzstyle{every state}=[fill=red,draw=none,text=white]
    
      \node (E)                     {$s_4\cdot s_3\equiv x\cdot\cos(\log x)$};
      \node (D) [below right of=E]  {$s_3\equiv\cos(s_2)$};
      \node (C) [below of=D]        {$s_2\equiv\log(s_1)$};
      \node (B) [below of=C]        {$s_1\equiv x$};
      \node (A) [below left of=E]   {$s_4\equiv x$};
    
      \path (A) edge node {} (E)
            (B) edge node {} (C)
            (C) edge node {} (D)
            (D) edge node {} (E);
    \end{tikzpicture}
\end{center}\captionof{figure}{\centering Computational graph of the algorithmic differentiation of the function $f(x) = x\cdot\cos(\log x)$.}

\tab\\Based on this graph, let us define the quantum states to calculate the derivative of this function.
\begin{itemize}
    \item $\ket{s_1} = \ket{x_0}\otimes\ket{0}\otimes\ket{1} \equiv \ket{val_1}\otimes\ket{0}\otimes\ket{der_1}$ 
    \item $\ket{s_2} = \texttt{ADlog}(\ket{s_1})$
    \item $\ket{s_3} = \texttt{ADcos}(\ket{s_2})$
    \item $\ket{s_4} = \texttt{ADtimes}(\ket{s_4},\ket{s_3})$
\end{itemize}

Let us now explicitly calculate the \textit{valder} states which will lead to the derivative of the function $f(x) = x\cdot\cos(\log x)$. Using the definition of \texttt{ADlog} we have for $\ket{s_2}$:
\begin{eqnarray*}
    \ket{s_2}  & = & \texttt{ADlog}(\ket{s_1}) \\
                & = & \ket{\log val_1} \otimes \ket{0} \otimes \ket{\frac{1}{val_1}\cdot der_1} \\
                & = & \ket{\log x} \otimes \ket{0} \otimes \ket{\frac{1}{x_0}\cdot 1} \\
                & \equiv & \ket{val_2} \otimes \ket{0} \otimes \ket{der_2}
\end{eqnarray*}

For $\ket{s_3}$, using the definition of \texttt{ADcos}:
\begin{eqnarray*}
    \ket{s_3}  & = & \texttt{ADcos}(\ket{s_2}) \\
                & = & \ket{\cos (val_2)} \otimes \ket{0} \otimes \ket{-\sin(val_2)\cdot der_2} \\
                & = & \ket{\cos (\log x_0)} \otimes \ket{0} \otimes \ket{-\sin(\log x_0)\cdot\frac{1}{x_0}} \\                
                & \equiv & \ket{val_3} \otimes \ket{0} \otimes \ket{der_3}
\end{eqnarray*}

Finally, using the definition of \texttt{ADtimes} for the states $\ket{s_3}$ and $\ket{s_4}$ we get:
\begin{eqnarray*}
    \texttt{qAD}\left[x\cdot\cos(\log x)\right]  
                & \equiv & \ket{f^\prime} = \texttt{ADtimes}(\ket{s_4},\ket{s_3}) \\
                & = & \ket{val_4\cdot val_3} \otimes \ket{0} \otimes \ket{der_4\cdot val_3+val_4\cdot der_3} \\
                & = & \ket{x\cdot\cos(\log x_0)} \otimes \ket{0} \otimes \ket{1\cdot\cos(\log x_0)+x_0\cdot\left(-\sin(\log x_0)\cdot\frac{1}{x_0}\right)} \\
                & = & \ket{x_0\cdot\cos(\log x_0)} \otimes \ket{0} \otimes \ket{\cos(\log x_0)-\sin(\log x_0)}
\end{eqnarray*}

The state $\ket{f^\prime}$ contains then the value of the function and of its derivative, both valuated at point $x_0$.

\section{A note on error and cost estimation}\label{sec-note-error}

The algorithms underlying the qAD for the calculation of the primitive
functions \cite{2015arXiv151108253B} have been developed within a
fixed-precision framework. In order to apply these primitive functions in the
qAD algorithm, we assume that the calculations outlined until now are also
performed with a fixed-point arithmetic. This choice naturally leads to
a possible precision loss error. Precision loss is related to
the fact that for calculations other than a simple summation the number of
fractional digits might be larger than the fractional digits of the input
values due to the accumulation of truncation error. 

The error can be calculated starting from the
computational graph of the function under consideration, which has a node for
each input and intermediate operation (primitive or arithmetic expression) and
a directed edge representing the value/derivative (valder) to be used. In
particular, if $r$ is the number of the nodes of the computational graph, at
each step we define an intermediate \textit{valder} variable $(v_{k},d_{k})$
in which both value and derivative are truncated to a number of bits on the right of the decimal point, say $b$, and passed over to the next node. The error will depend on
the specific component (primitive function or arithmetic expression) which is
being calculated. For instance, the error on the inverse function, $f(x) =
\frac1 x$, is bounded by a value $\frac{2+\log_{2}b}{2^{b}}$
\cite{2015arXiv151108253B}.

For a single-variable composite function $f(x)$, with a computational graph
with $r+1$ nodes, let us denote with $\epsilon_{k}$ the error of the
evaluation of the function and with $\delta_{k}$ the error of the evaluation
of the derivative at $k$-th node. These errors are the result of the
truncation of the numbers due to the limited number of qubits,
say $b$, of the fractional part. Please note that these errors depend on the
specific primitive function or arithmetic expression at a given node. Thus, at
each node the value of the function accumulates an error:
\[
v_{k}\rightarrow\hat{v}_{k}=v_{k}+\epsilon_{k}%
\]
and the derivatives
\[
d_{k}\rightarrow\hat{d}_{k}=d_{k}+\delta_{k}.
\]
For $r=0$ the first step corresponds to $v_{0}=x_{0}$, but the corresponding
ket contains the number $\hat{v}_{0}$ with the corresponding discrepancy
$\epsilon_{0}.$ The inital derivative is $\hat{d}_{0}=d_{0}=1$ (no error).

If, for instance, we study the derivative of the function $f(g(x))$ the node
$r=1$ is given by
\begin{equation}
\hat{v}_{1}=g(\hat{v}_{0})+\epsilon_{1}=g(x_{0}+\epsilon_{0})+\epsilon_{1}%
\end{equation}
and%
\begin{equation}
\hat{d}_{1}=d_{1}+\delta_{1}=g^{\prime}(x_{0}+\epsilon_{0})+\delta_{1}%
\end{equation}
where $\delta_{1}$ is the error associated to the function $g^{\prime}(x).$
The next and final node ($r=2$) is realized for%
\begin{equation}
\hat{v}_{2}=f(g(x_{0}+\epsilon_{0})+\epsilon_{1})+\epsilon_{2}%
\end{equation}
and
\begin{equation}
\hat{d}_{2}=d_{2}+\delta_{2}=f^{\prime}(g(x_{0}+\epsilon_{0})+\epsilon
_{1})\left[  g^{\prime}(x_{0}+\epsilon_{0})+\delta_{1}\right]  +\delta
_{2}\text{,}%
\end{equation}
where $\delta_{2}$ is the error associated to $f^{\prime}(x).$ 

It is useful to define the function
\begin{equation}
F(x,\epsilon_{0},\epsilon_{1},\epsilon_{2})=f(g(x+\epsilon_{0})+\epsilon
_{1})+\epsilon_{2}%
\end{equation}
such that $F(x,0,0,0)=f(g(x)).$ Then, an upper bound on the total error on the
function evaluation can be estimated as follows:%
\[
\left\vert F(x,\epsilon_{0},\epsilon_{1},\epsilon_{2})-F(x,0,0,0)\right\vert
\approx\left\vert \sum_{k=0}^{2}\left(  \frac{\partial F}{\partial\epsilon
_{k}}\right)  _{\epsilon_{k}=0}\cdot\epsilon_{k}\right\vert \leq\sum_{k=0}%
^{2}\left\vert \left(  \frac{\partial F}{\partial\epsilon_{k}}\right)
_{\epsilon_{k}=0}\right\vert \cdot\left\vert \epsilon_{k}\right\vert \text{ .}%
\]
For the estimation of the error on the derivative, we define the function
\begin{equation}
G(x,\epsilon_{0},\epsilon_{1},\delta_{1},\delta_{2})=f^{\prime}(g(x+\epsilon
_{0})+\epsilon_{1})\left[  g^{\prime}(x+\epsilon_{0})+\delta_{1}\right]
+\delta_{2}\text{ ,}%
\end{equation}
such that $G(x,0,0,0,0)=f^{\prime}(g(x))g^{\prime}(x).$ Then, the upper bound of the total error of the derivative (the main quantity that we calculate) is given by:
\begin{equation}
\left\vert G(x,\epsilon_{0},\epsilon_{1},\delta_{1},\delta_{2}%
)-G(x,0,0,0,0)\right\vert \approx\nonumber
\end{equation}
\begin{equation}
\approx\left\vert \sum_{k=0}^{1}\left(
\frac{\partial G}{\partial\epsilon_{k}}\right)  _{\epsilon_{k}=0,\delta_{k}%
=0}\cdot\epsilon_{k}+\sum_{k=1}^{2}\left(  \frac{\partial G}{\partial
\delta_{k}}\right)  _{\epsilon_{k}=0,\delta_{k}=0}\cdot\delta_{k}\right\vert\nonumber
\end{equation}%
\begin{equation}
\leq\sum_{k=0}^{1}\left\vert \left(  \frac{\partial G}{\partial\epsilon_{k}%
}\right)  _{\epsilon_{k}=0,\delta_{k}=0}\cdot\epsilon_{k}\right\vert
+\sum_{k=1}^{2}\left\vert \left(  \frac{\partial G}{\partial\delta_{k}%
}\right)  _{\epsilon_{k}=0,\delta_{k}=0}\cdot\delta_{k}\right\vert \text{ .}%
\end{equation}

This procedure can be generalized to an arbitrary composite function (including
arithmetic operations) involving $r+1$ nodes, leading to the introduction of
analogous functions $F(x,\epsilon_{0},...,\epsilon
_{r})$ and  $G(x,\epsilon_{0},...,\epsilon
_{r-1},\delta_{1},...,\delta_{r}).$ Hence, the upper bounds on the function
and on its derivative are: %
\begin{equation}
\left\vert F(x,\epsilon_{0},\epsilon_{1},\epsilon_{2},...,\epsilon
_{r})-F(x,0,0,...,0)\right\vert \leq\sum_{k=0}^{r}\left\vert \left(
\frac{\partial F}{\partial\epsilon_{k}}\right)  _{\epsilon_{k}=0}\cdot
\epsilon_{k}\right\vert
\end{equation}
\begin{equation}
\left\vert G(x,\epsilon_{0},\epsilon_{1},...,\epsilon_{r-1},\delta
_{1},...,\delta_{r})-G(x,0,...,0)\right\vert\leq\nonumber
\end{equation}
\begin{equation}
\leq\sum_{k=0}^{r-1}\left\vert
\left(  \frac{\partial G}{\partial\epsilon_{k}}\right)  _{\epsilon_{k}%
=\delta_{k}=0}\cdot\epsilon_{k}\right\vert +\sum_{k=1}^{r}\left\vert \left(
\frac{\partial G}{\partial\delta_{k}}\right)  _{\epsilon_{k}=\delta_{k}%
=0}\cdot\delta_{k}\right\vert \text{ }%
\end{equation}

The cost of the algorithm fully depends on steps 3(a)-3(d) in the
qAD algorithm. These steps have a computational complexity depending on the
implementation of the specific component (or node) of the computational graph
under consideration. 

The cost of qAD algorithm can be computed by adding the costs of the individual algorithms evaluating the functions in the computational graph for Algorithmic
Differentiation. For the functions presented in section \ref{section-qad-framework} these costs are known \cite{2013NJPh...15a3021C, 2015arXiv151108253B, 2018arXiv180503265H}. 

Alternatively, and for simplicity, we may assume that the number of operations
required by the elementary function and derivative evaluations is
uniformly bounded by a quantity $c$. Thus, if the number of nodes in
the computational graph is $r$ then the overall cost of the algorithm is
bounded from above by a quantity proportional to $r \cdot c$.

\section{Conclusions}
\label{sec-conclusions}
This paper presents an extension of the algorithmic differentiation procedure for a classical computer to a quantum computer and shows a framework for implementing quantum Algorithmic Differentiation for scientific computing.

We first defined the \textit{valder} quantum state which is the input state to calculate the value of a function and its derivative in a given point. The \textit{valder} state is in fact a quantum state representation of the point at which we need to calculate the function and its derivative, say $x_0$. This state is composed by three quantum registers: the first representing the value $x_0$, the second representing the value $0$, and finally the third representing the value $1$. The latter two states are needed to calculate the value of the derivative of the function in the quantum algorithmic differentiation framework. 

We then defined the operators and procedures necessary to calculate the function and the derivative in $x_0$. The \textit{transfer} or \textit{copy} operators copies the first quantum register onto the second, the \textit{reset} procedure sets the second register to represent the value $0$ and finally the AD operator is built in order to compute the value of a primitive function and its derivative. We also showed two possible solutions for the reset procedure, one fully quantum and the other applicable in hybrid systems, with the help of a classical bit.

In conclusion, based on the existing definitions of primitive functions on a quantum computer, we outlined the algorithm to compute the value of a composite function and its derivative at a given point, based on the computational graph of the function itself.

We acknowledge the effect on the cost of the current implementation of this algorithm due to the large number of qubits needed to represent the value of functions and derivatives in a point. 
On the other hand, the use of a very small number of qubits might not adequately satisfy the error constraints for the task at hand. There are thus open questions regarding the actual implementation of the algorithm and the issues related to the trade-off of precision and efficiency that need to be addressed as follow up of the current paper.  

Nevertheless, we do see a potential in applying our quantum algorithmic
differentiation framework to existing scientific computing, optimization
and machine learning algorithms.

\nonumsection{Acknowledgements}
\noindent
GC thanks Limor Arieli, Marko Iskra and Arturo De Marinis for the support and useful discussions.
The authors also thank the unknown referee for the valuable suggestions on the error analysis and the numerical framework of the algorithm.

\bibliography{bibliography}{}
\bibliographystyle{unsrt}

\end{document}